\documentclass[preprint2]{aastex}
\usepackage{natbib}
\shorttitle{Radial Stellar Pulsation and 3D Convection. I.}
\shortauthors{Geroux and Deupree}

\begin{document}

\title{RADIAL STELLAR PULSATION AND 3D CONVECTION. I. NUMERICAL METHODS AND ADIABATIC TEST CASES }

\author{Chris M. Geroux\altaffilmark{1}}
\affil{Department of Astronomy \& Physics, Saint Mary's University, Halifax, NS}
\email{cgeroux@ap.smu.ca}
\author{Robert G. Deupree\altaffilmark{1}}
\affil{Department of Astronomy \& Physics, Saint Mary's University, Halifax, NS}

\altaffiltext{1}{Institute for Computational Astrophysics}

\begin{abstract}
We are developing a 3D radiation hydrodynamics code to simulate the interaction of convection and pulsation in classical variable stars. One key goal is the ability to carry these simulations to full amplitude in order to compare them with observed light and velocity curves. Previous 2D calculations were prevented from doing this because of drift in the radial coordinate system, due to the algorithm defining radial movement of the coordinate system during the pulsation cycle. We remove this difficulty by defining our coordinate system flow algorithm to require that the mass in a spherical shell remain constant throughout the pulsation cycle. We perform adiabatic test calculations to show that large amplitude solutions repeat over more than 150 pulsation periods. We also verify that the computational method conserves the peak kinetic energy per period, as must be true for adiabatic pulsation models.
\end{abstract}


\section{INTRODUCTION}
Early nonlinear calculations of stellar pulsation, as outlined by \cite{Christy-1964}, used a 1D Lagrangian framework and had considerable success producing full amplitude RR Lyrae models that resembled the basic observations. However, these calculations used purely radiative envelopes and failed to identify a red edge to the RR Lyrae instability strip. This lead to the hypothesis that convection in the ionization zones quenched pulsation \citep{Baker-1965,Christy-1966b}. To explore this, \cite{Tuggle-1973} used a 1D, linear, non-adiabatic code with a time-independent mixing length theory for the convective flux. They found that time-independent convection only reduced the growth rate of pulsation but did not produce pulsational stability.

Several formalisms and 1D codes were developed that included time-dependent convection by introducing an additional differential equation to calculate the evolution of the convective flux with time based on the standard mixing length theory \citep{Cox-1966a,Cox-1966b,Unno-1967,Gough-1977}. \cite{Stellingwerf-1982a, Stellingwerf-1982b, Stellingwerf-1984a, Stellingwerf-1984b, Stellingwerf-1984c} developed a time-dependent treatment of convection for 1D Lagrangian models which follows the time evolution of averaged convective velocities and includes a treatment of overshooting and eddy viscosity to account for small length scale kinetic energy dissipation.

There are several numerical difficulties associated with modeling radial pulsation. First there are very steep gradients in the ionization zones and adequately resolving these gradients is important for accurate modeling. \cite{Gehmeyr-1992a, Gehmeyr-1992b,Gehmeyr-1993} and \cite{Dorfi-1991} have both included adaptive grids too better resolve the steep gradients in the ionization zones as they sweep through the envelope during pulsation. Using a version of Stellingwerf's time-dependent convective model with his adaptive scheme, \citeauthor{Gehmeyr-1992a} was able to produce a red edge at roughly the observed effective temperature. He notes that the effective temperature of the red edge is dependent on the parameters used for the convective model, and that the predicted temperature of the red edge could vary by a few hundred degrees Kelvin depending on the values used for the convective model parameters. Also, there are differences between Gehmeyr's light amplitude-rise time relationship and the observed relationship in both slope and zero point. \cite{Dorfi-1996} used their adaptive code to calculate light and radial velocity curves which exhibit shapes typical of RR Lyrae stars. A second potential difficulty is an accurate representation of the surface boundary, which \citeauthor{Dorfi-1996} tested by including a model atmosphere and found that its inclusion did not impact the pulsational characteristics of the model.

\cite{Marconi-2003} used the 1D, Lagrangian, hydrodynamics code described by \cite{Bono-1994} and \cite{Bono-1997a,Bono-1997b} to compute RR Lyrae models to compare with the RR Lyrae stars observed in M3. In order to fully specify the problem \citeauthor{Marconi-2003} needed to choose a mixing-length parameter, and adopted both $l/H_p=1.5$ and $2.0$, where $l$ is the mixing-length and $H_p$ is the pressure scale height. They found that in order to match the boundaries of RR Lyrae gap in M3, they required two different mixing-length parameters, one to obtain the observed blue edge location ($l/H_p\approx 1.5$) and the other to obtain the observed red edge location ($l/H_p\approx 2.0$). In addition the observed visual amplitude as a function of B-V displays nonlinear characteristics, while theoretical relations predict linear relationships. \citeauthor{Marconi-2003} also mention that a mixing-length parameter of 2.0 produces luminosities for horizontal-branch models that are brighter than what is observed by $\approx0.08\pm0.05$ mag. 

Other models for time-dependent convection in one dimension have been proposed by \cite{Kuhfuss-1986} and \cite{Xiong-1989}. \citeauthor{Kuhfuss-1986} argued that the convective model by \cite{Stellingwerf-1982a} does not use the diffusion approximation consistently throughout the model. \cite{Smolec-2008} developed their application of the \citeauthor{Kuhfuss-1986} convective model, which requires eight free parameters, and used it to study convection in $\beta$ Cephei stars \citep{Smolec-2007}. They found that convection is not important for calculating pulsation amplitudes for their models. However, they caution that their convective model, while working well for classical pulsators, is at the limits of its applicability in the $\beta$-Cephei models they are studying. More recently, \cite{Olivier-2005} have also developed a code including the \citeauthor{Kuhfuss-1986} convective model and present some test calculations of their program, mentioning that the turbulent viscosity parameter shows potential as an important determinant of the pulsation amplitudes.

The distillation of multi-dimensional convective phenomenon to one dimension is always accompanied by extra equations and/or parameters to approximate the effects of convective motions of material in more than one spatial dimension. \cite{Deupree-1977a} approached the interaction of convection and stellar pulsation in a fundamentally different way using a two-dimensional hydrodynamic code directly following the convective flow patterns. While \cite{Deupree-1977b,Deupree-1977c,Deupree-1980,Deupree-1985} was able to successfully determine the observed edges of the RR Lyrae instability strip, he was unable to compute full amplitude solutions because his algorithm for moving the radial coordinate allowed the radial zoning to drift over time. Consequently at later times, the radial zoning did not cover the hydrogen ionization zone adequately and the calculations were eventually numerically unreliable. 
The algorithm \citeauthor{Deupree-1977a} used for the moving radial coordinate used the horizontal average of the radial velocities at a particular radius as the grid velocity. Recently \cite{Bruenn-2010} have used a similar average radial velocity as the grid velocity to follow the core in fall phase in supernovae simulations. Another form of a radial moving grid was by \cite{Mundprecht-2009}, where the radial grid velocity at the surface was set as the average of the radial velocities, and the inner grid velocity was held constant. The intermediate radial grid velocities were then set using a dilatation factor.

Recently \cite{Stokl-2008} developed an approach for model pulsation somewhere between the 2D model of \citeauthor{Deupree-1977a} and the 1D convective models. \citeauthor{Stokl-2008} used two radial columns to model convection. One column represented the sum of all upward convective flows, and the other column represented the sum of all downward flows. While not including any mixing length parameter, it does contain free parameters related to the physical size of the convective cells, and the fraction of the surface area of a spherical surface which contains downdrafts. These parameters do have a physical basis, but in practice it would be difficult to determine them as they are probably functions of depth and likely depend on a particular star's properties. Also, because the model only has two radial columns it may miss some of the more subtle features of convection. More recently others have begun working on directly simulating the interaction of convection and pulsation in 2D \citep{Muthsam-2010b, Gastine-2010}.

Both \cite{Buchler-2009} and \cite{Marconi-2009} have stressed the importance of improving the convective models used in variable stars. \citeauthor{Buchler-2009} highlights some of the well known difficulties facing time-dependent mixing length noting that it is an empirical description, rather than a consistent physical description. Also, the up to eight or more free parameters used in time-dependent mixing length approach can not be chosen based on physics, but instead must be calibrated by comparison with observations. \citeauthor{Marconi-2009} mentions some remaining problems for RR Lyrae models, particularly the unsatisfactory match between theoretical light curve morphology and observed light curve morphology near the red edge of the RR Lyrae instability strip, supporting the suggestion that an improved treatment of turbulent convection is needed.

3D convective simulations have had significant success in other areas of stellar astrophysics. For example \cite{Nordlund-2009} note many of the recent successes in 3D modeling of solar surface convection, in particular that 3D models with numerical resolutions around $200^3$ reproduce widths, shifts, and shapes of observed photospheric spectral lines with high accuracy. 3D convective simulations by \cite{Meakin-2007} simulated core convection in a massive star finding differences in 2D and 3D convective velocities in both morphology and magnitude. 3D models remove the need for many free parameters and the lack of a physically consistent description of convection, with convection resulting naturally from the conservation laws; however, an algorithm to include the effects of sub-grid scale turbulence on the larger eddies is still required. Here we build on the ideas of \cite{Deupree-1977a}, with the goal of developing a fully 3D calculation in which the radial coordinate moves in such a manner to allow us to perform full amplitude solutions of RR Lyrae models with Large Eddy Simulations (LES) of convective energy transport. As a first step we apply our approach to purely adiabatic models to verify that the method can compute accurate and stable large amplitude periodic solutions over many periods.

\section{HORIZONTAL EULERIAN RADIAL LAGRANGIAN SCHEME}
\label{sec:HERLS}
Calculation of full amplitude solutions requires following the pulsation for many periods. 1D codes have been able to calculate full amplitude solutions, while \citeauthor{Deupree-1977a}'s 2D code had difficulty after many periods because of his particular moving radial coordinate. This led us to try using the internal mass, $M_r$, as the radial independent variable instead of radius, $r$, and allow $r$ to change such that the mass within a shell remains constant. This requires introducing a grid velocity, $v_{0r}$, in the radial direction that dictates how the coordinate system radius changes. The intent is that our radial grid acts like that of a 1D Lagrangian code while allowing the usual Eulerian approach in the horizontal directions. Note that this approach still allows fluid flow across radial zone boundaries. It just moves the radial gridding so that it maintains the mass in a spherical shell. It does not put any constraints on the horizontal flow and does not alter the physics of the conservation equations in any way.

The calculations are performed in a limited range of the spherical polar coordinates $\theta$ and $\phi$, a 3D version of a ``pie slice''. Periodic boundary conditions are placed in the horizontal directions. The interior boundary is placed at a location deeper than 0.15 of the stellar radius and is regarded as rigid. This is a common assumption in most 1D simulations. Because the horizontal zoning would get very narrow (leading, through the Courant condition, to undesirably short time steps) and because the 3D flow of interest is expected to be only in the surface ionization regions, we impose a purely radial region for an arbitrary number of radial zones above the interior boundary. In conjunction with the assumption that non-radial motion occurs only near the very low mass surface, we assume the gravitational force has its spherically symmetric form.

\subsection{Conservation Equations}
\label{sec:cons-eqs}
We first define a horizontally averaged density which allows us to replace an infinitesimal radial change, $dr$, with an infinitesimal internal mass change, $dM_r$. The mass of a spherical shell of thickness ${\rm d}r$ is given by ${\rm d}M_r=4\pi r^2\langle\rho\rangle {\rm d} r$ where $\langle\rho\rangle$ is the volume averaged density within a spherical shell,
\begin{equation}
\label{eq:rho-ave}
\langle\rho\rangle_i=\frac{1}{V_i}\sum_{j,k}\rho_{i,j,k}V_{i,j,k}.
\end{equation}
Here $i$, $j$, and $k$ are the $\hat{r}$, $\hat{\theta}$ and $\hat{\phi}$ zone indices defined at zone centers and increase with each of their respective spherical coordinates (e.g., $i$ increases from the center of the star towards the surface). $V_i$ is the volume of a spherical shell at a particular radial shell, $i$, that spans all the $j$ and $k$ at that $i$. $V_{i,j,k}$ is the volume of the ($i$,$j$,$k$) grid cell. In order to develop and test this approach, we have assumed that the system is adiabatic. Introducing both $dM_r$ and the grid velocity, $v_{0r}$, into the 3D conservation equations for mass, three components of momentum, and energy produces:
\begin{eqnarray}
\label{eq:mass-cons-fd}
\frac{\partial \rho}{\partial t}&+&4\pi r^2\langle\rho\rangle\left(v_r-v_{0r}\right)\frac{\partial \rho}{\partial M_r}+\frac{v_\theta}{r}\frac{\partial \rho}{\partial\theta}\nonumber \\
{}&+&\frac{v_\phi}{r\sin\theta}\frac{\partial\rho}{\partial\phi}{}+4\pi\langle\rho\rangle\rho\frac{\partial \left(r^2v_r\right)}{\partial M_r}\nonumber \\
{}&+&\frac{\rho}{r\sin\theta}\frac{\partial \left(v_\theta\sin\theta \right)}{\partial\theta}{}+\frac{\rho}{r\sin\theta}\frac{\partial v_\phi}{\partial \phi}\nonumber \\
{}&=&0,
\end{eqnarray}
\begin{eqnarray}
\label{eq:rad-mom-cons}
\frac{\partial v_r}{\partial t}&+&4\pi r^2\langle\rho\rangle\left(v_r-v_{0r}\right)\frac{\partial v_r}{\partial M_r}+\frac{v_\theta}{r}\frac{\partial v_r}{\partial \theta}\nonumber \\
&+&\frac{v_\phi}{r\sin\theta}\frac{\partial v_r}{\partial\phi}=\frac{-4\pi r^2\langle\rho\rangle}{\rho}\frac{\partial P}{\partial M_r}\nonumber \\
&+&\frac{v_\theta^2}{r}+\frac{v_\phi^2}{r}-\frac{GM_r}{r^2},
\end{eqnarray}
\begin{eqnarray}
\label{eq:theta-mom-cons}
\frac{\partial v_\theta}{\partial t}&+&4\pi r^2\langle\rho\rangle\left(v_r-v_{0r}\right)\frac{\partial v_\theta}{\partial M_r}+\frac{v_\theta}{r}\frac{\partial v_\theta}{\partial \theta}\nonumber \\
&+&\frac{v_\phi}{r\sin\theta}\frac{\partial v_\theta}{\partial\phi}=\frac{-1}{r\rho}\frac{\partial P}{\partial \theta}+\frac{v_\phi^2\cot\theta}{r}\nonumber \\
&-&\frac{v_r v_\theta}{r},
\end{eqnarray}
\begin{eqnarray}
\label{eq:phi-mom-cons}
\frac{\partial v_\phi}{\partial t}&+&4\pi r^2\langle\rho\rangle\left(v_r-v_{0r}\right)\frac{\partial v_\phi}{\partial M_r}+\frac{v_\theta}{r}\frac{\partial v_\phi}{\partial \theta}\nonumber \\ 
&+&\frac{v_\phi}{r\sin\theta}\frac{\partial v_\phi}{\partial\phi}=\frac{-1}{\rho r\sin\theta}\frac{\partial P}{\partial \phi}-\frac{v_rv_\phi}{r}\nonumber \\
&-&\frac{v_\theta v_\phi \cot\theta}{r},
\end{eqnarray}
\begin{eqnarray}
\label{eq:E-cons}
\frac{\partial E}{\partial t}&+&4\pi r^2\langle\rho\rangle\left(v_r-v_{0r}\right)\frac{\partial E}{\partial M_r}+\frac{v_\theta}{r}\frac{\partial E}{\partial\theta}\nonumber \\
&+&\frac{v_\phi}{r\sin\theta}\frac{\partial E}{\partial\phi}+\frac{4\pi\langle\rho\rangle P}{\rho }\frac{\partial \left(r^2v_r\right)}{\partial M_r}\nonumber \\
&+&\frac{P}{\rho r\sin\theta}\frac{\partial \left(v_\theta\sin\theta \right)}{\partial \theta}+\frac{P}{\rho r\sin\theta}\frac{\partial v_\phi}{\partial \phi}\nonumber \\
&=&0.
\end{eqnarray}
The above symbols have their usual meaning ($E$ is the specific internal energy). Finally the system is closed with three further equations:
\begin{equation}
\label{eq:r-update}
\frac{\partial r}{\partial t}=v_{0r},
\end{equation}
which is used to update the radius, 
\begin{equation}
\label{eq:eos}
P=(\gamma-1)\rho E,
\end{equation}
 a simple gamma-law gas for the equation of state, and an equation for solving for the grid velocity (see equation \ref{eq:rad-grid-vel}). Before we present the equation for the grid velocity, we present an equivalent equation to equation (\ref{eq:mass-cons-fd}) by which we solve mass conservation.

\subsection{Finite Volume Mass Conservation}
\begin{figure}[tb!]
\center
\plotone{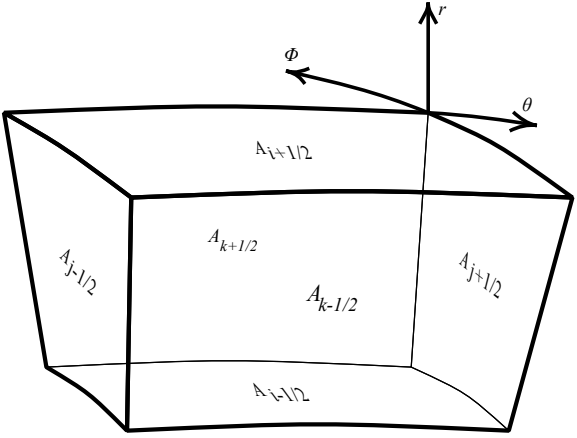}
\caption{ Geometry of a cell in spherical coordinates. The areas of the six surfaces of the ``cube'' are shown.}
\label{fig:cell}
\end{figure}
\label{sec:fv-mass-cons}
The definition of volume we will use in the equation for determining the grid velocity (eq. [\ref{eq:rad-grid-vel}]) and the average density (eq. [\ref{eq:rho-ave}]) must be consistent with the mass conservation equation. The definition of volume we use in equations (\ref{eq:rad-grid-vel}) and (\ref{eq:rho-ave}) is
\begin{equation}
V_{i,j,k}=\int_{r_{i-1/2}}^{r_{i+1/2}}\!\!\!\!\int_{\theta_{j-1/2}}^{\theta_{j+1/2}}\!\!\!\!\int_{\phi_{i-1/2}}^{\phi_{j+1/2}}\!\!\!r^2\sin\theta \, d\phi\, d\theta\, dr.
\end{equation}
To make the mass equation consistent with this definition of the volume, one integrates equation (\ref{eq:mass-cons-fd}) over the volume, then uses Gauss's theorem to convert the volume integral into a surface integral, producing the finite volume form of the mass conservation equation.

The mass in a cell changes only from mass flowing into and out of that cell. This change from one time step ($n$) to the next ($n+1$) can be written as
\begin{eqnarray}
\label{eq:mass-cons-fv}
&{}&V_{i,j,k}^{n+1}\rho_{i,j,k}^{n+1}-V_{i,j,k}^n\rho_{i,j,k}^n=\Delta t^{n+1/2}\nonumber \\
&\cdot&\left[ \left(F_{i-1/2}^{n+1/2}A_{i-1/2}^{n+1/2}-F_{i+1/2}^{n+1/2}A_{i+1/2}^{n+1/2}\right)\right.\nonumber \\
&+&\left(F_{j-1/2}^{n+1/2}A_{j-1/2}^{n+1/2}-F_{j+1/2}^{n+1/2}A_{j+1/2}^{n+1/2}\right)\nonumber \\
&+&\left.\left(F_{k-1/2}^{n+1/2}A_{k-1/2}^{n+1/2}-F_{k+1/2}^{n+1/2}A_{k+1/2}^{n+1/2}\right)\right],
\end{eqnarray}
where $A$ is the area of a cell face (see figure \ref{fig:cell}) and the $F$s are fluxes defined as
\begin{eqnarray}
F_{i\pm 1/2}^{n+1/2}&=&\left(v_{r,i\pm 1/2,j,k}^{n+1/2}\right.-\left. v_{0r,i\pm 1/2}^{n+1/2}\right)\rho_{i\pm 1/2,j,k}^{n},\label{eq:outer-flux}\\
F_{j\pm 1/2}^{n}&=&v_{\theta,i,j\pm 1/2,k}^{n+1/2}\rho_{i,j\pm 1/2,k}^{n},\\
F_{k\pm 1/2}^{n+1/2}&=&v_{\phi,i,j,k\pm 1/2}^{n+1/2}\rho_{i,j,k\pm 1/2}^{n}.
\end{eqnarray}
To obtain the densities at the interfaces ($\rho_{i\pm 1/2}^n$) straight averages are computed of centered densities adjacent to the interface. The two centered subscripts have intentionally been omitted in the expressions for the fluxes and areas to reduce equation length (e.g. $i$ was left off but $i+1/2$ was kept). Note that the densities in the fluxes are at time $n$ and not $n+1/2$, so our solution algorithm is straightforwardly explicit, as is true for many 1D calculations. It is not possible to properly time center all terms without introducing a more complex implicit or multi-step explicit algorithm. With this expression for the fluxes, we can directly solve equation (\ref{eq:mass-cons-fv}) for the density at the new time step.

\subsection{Radial Grid Velocity}
\label{sec:rad-grid-vel}
The final piece needed to complete the description is the calculation of the grid velocity. For a spherical shell to have constant mass, the net flow of mass into and out of that spherical shell must be zero.  Summing up all the fluxes into and out of the individual horizontal cells in a spherical shell, substituting equation (\ref{eq:outer-flux}) in for the outer radial flux (at $i+1/2$) and setting the result equal to zero we arrive at the equation,
\begin{eqnarray}
0&=&\sum_{jk}\left[\left(F_{i-1/2}^{n+1/2}A_{i-1/2}^{n+1/2}\right.\right.\nonumber \\
{}&-&\left.\left(v_{r,i+1/2}^{n+1/2}-v_{0r,i+1/2}^{n+1/2}\right)\rho_{i+1/2}^{n}A_{i+1/2}^{n+1/2}\right)\nonumber \\
{}&+&\left(F_{j-1/2}^{n+1/2}A_{j-1/2}^{n+1/2}-F_{j+1/2}^{n+1/2}A_{j+1/2}^{n+1/2}\right)\nonumber \\
{}&+&\left.\left(F_{k-1/2}^{n+1/2}A_{k-1/2}^{n+1/2}-F_{k+1/2}^{n+1/2}A_{k+1/2}^{n+1/2}\right)\right].
\end{eqnarray}
Solving for the outer grid velocity, $v_{0r,i+1/2}^{n+1/2}$, produces an equation for calculating the new grid velocity, 
\begin{eqnarray}
\label{eq:rad-grid-vel}
v_{0r,i+1/2}^{n+1/2}&=&\frac{-1}{\sum_{j,k}A_{i+1/2}^{n+1/2}\rho_{i+1/2,j,k}^{n}}\nonumber \\
{}&\cdot&\sum_{jk}\left[\left(F_{i-1/2}^{n+1/2}A_{i-1/2}^{n+1/2}-v_{r,i+1/2}^{n+1/2}\rho_{i+1/2}^{n}A_{i+1/2}^{n+1/2}\right)\right.\nonumber \\
{}&+&\left(F_{j-1/2}^{n+1/2}A_{j-1/2}^{n+1/2}-F_{j+1/2}^{n+1/2}A_{j+1/2}^{n+1/2}\right)\nonumber \\
{}&+&\left.\left(F_{k-1/2}^{n+1/2}A_{k-1/2}^{n+1/2}-F_{k+1/2}^{n+1/2}A_{k+1/2}^{n+1/2}\right)\right].
\end{eqnarray}

The inner radial flux, $F_{i-1/2}^{n+1/2}$, is dependent on the grid velocity at the inner interface. At the first radial zone boundary next to the rigid core we impose both a zero radial velocity and grid velocity. Thus, equation (\ref{eq:rad-grid-vel}) can be solved recursively from the model interior boundary to the surface to determine the grid velocity at all interfaces.

\section{COMPUTATIONAL SETUP}

\subsection{Starting Model}
The initial model for our adiabatic simulations is generated by requiring that it be in hydrostatic equilibrium. When this constraint is applied to the conservation equations the only terms that remain are the pressure and gravity terms in the radial momentum conservation equation. In particular, there are no terms left in the internal energy conservation equation, and thus no equation to solve for the energy structure. To provide this information, an energy profile was generated from another stellar modeling code ROTORC \citep{Deupree-1990} and energies were interpolated in $\log(M_r)$ to cell centers. Once we impose the energy distribution, we can simultaneously solve the radial hydrostatic equilibrium finite difference equation and the equation of state for the pressure and density structure of the model given the spacing of the independent variable $M_r$. The radius is determined from the volume required to produce the calculated density from the mass of the shell. No convective model is included in the starting model because RR Lyrae do not have extensive convective regions to affect the structure. To induce pulsation a radial velocity profile from the linear, non-adiabatic, radial pulsation code, LNA, \citep{Castor-1971}, modified to allow a gamma law gas, is imposed so that the model pulsates around the equilibrium point in either the fundamental or the first overtone modes.

\subsection{The Grid and Numerical Details}
The simulation volume is separated into cells bounded by intersecting surfaces. These surfaces are defined at constant values of the three independent variables $M_r$, $\theta$, and $\phi$. Dependent quantities $\rho$, $E$, and $P$ are defined at cell centers and dependent quantities $r$, $v_r$, $v_{0r}$, $v_\theta$, and $v_\phi$ are defined at appropriate cell interfaces. The models used for testing have 107 radial, 10 theta, and 10 phi zones. The inner 10 radial zones are handled in 1D as discussed in \S \ref{sec:HERLS}. The zone number at which the switch between 1D and 3D is made is chosen by the user. For the test cases used in this paper the total mass of the star was 0.575 M$_\sun$, with an initial mass spacing of $4.5\times 10^{-9}$ M$_\sun$ at the surface, and increasing by 10\% each shell into the star. Both the $\theta$ and $\phi$ zones have a spacing of 1$^\circ$, so that the total simulation volume covers 100 square degrees. 

The equations outlined in \S \ref{sec:cons-eqs} are in differential form and are approximated by appropriate finite difference expressions. Spatial differentials are approximated by differences between quantities at either cell centers or cell interfaces depending on whether the quantity being updated in time is interface centered or cell centered, respectively. Temporal differentials are approximated by differences between the current grid state and the updated grid state divided by the time step, $\Delta t$, computed as a fraction of the minimum time step allowed by the Courant condition for the model as a whole. This then allows us to explicitly solve for the updated grid state given the current grid state and the time step.

Equation (\ref{eq:mass-cons-fv}) is written in finite volume form with fluxes defined at cell faces. The velocities required for these fluxes are already interface centered; however, the densities are not. In general, quantities that are needed at interfaces but defined at cell centers, and quantities that are needed at cell centers but defined at interfaces are approximated by averages of adjacent quantities. We have used artificial viscosity given by \citep{von-Neumann-1950,Richtmyer-1967} to smooth out shocks with a threshold velocity of one-hundredth of the local sound speed for turning on the artificial viscosity and have used weighted donor cell to stabilize advection terms, with a weight of 0.1 on the upwind terms and 0.9 for centered terms.

\subsection{Order of Calculation}
The order of calculation follows that of \cite{Deupree-1977a} with a few minor modifications. We start by updating the three velocities using equations (\ref{eq:rad-mom-cons}, \ref{eq:theta-mom-cons}, and \ref{eq:phi-mom-cons}) from time $n-1/2$ to $n+1/2$ using quantities at $n$ ($\rho$, $\langle\rho\rangle$, $r$, and $P$) and quantities at $n-1/2$ ($v_r$, $v_{r0}$,$v_\theta$, and $v_\phi$). Next the grid velocity is calculated at time $n+1/2$ using equation (\ref{eq:rad-grid-vel}) working from inner boundary of the model to the surface in a recursive manner. The updated radius is computed with equation (\ref{eq:r-update}). The density is updated from $n$ to $n+1$ using the equation for mass conservation (eq. [\ref{eq:mass-cons-fv}]), with quantities at $n$ ($\rho$ and $r$), and quantities at $n+1/2$ ($v_r$, $v_{r0}$, $v_\theta$, and $v_\phi$). The energy is updated in a similar manner. The equation of state then allows us to compute the pressure at the new time step from the updated density and specific internal energy.

\subsection{Parallelism}
The code we have developed to perform these calculations has been named SPHERLS (Stellar Pulsation with a Horizontal Eulerian Radial Lagrangian Scheme). SPHERLS has been designed from the beginning to allow for parallel calculations using MPI protocols. The parallel design allows for domain decomposition in all three directions with the ability to vary the number of ghost cells (used to express the boundary conditions of the local domain) copied from other processors. Note that boundary conditions in this sense are not the global boundary conditions of the calculation but only the information required from other processors to be able to perform the calculations on the processor in question. Equations (\ref{eq:rad-mom-cons}), (\ref{eq:theta-mom-cons}), (\ref{eq:phi-mom-cons}), (\ref{eq:E-cons}), (\ref{eq:mass-cons-fv}) depend on only local quantities and are easily applied to the local grids on each processor. The equation to calculate the grid velocity (eq. [\ref{eq:rad-grid-vel}]) requires information across all $j$ and $k$ space. Using this equation with domain decomposition in the $j$ and $k$ directions would require additional message passing which has not yet been implemented, and thus currently limits domain decomposition to the radial direction only. This could change in the future and may become helpful for optimizing calculations for larger horizontal grids.

During program initialization, each processor can be assigned different equations to solve on their local grid, which allows one to divide the computational domain up into a 3D and a 1D region. The 1D region is composed of only a single zone at each $i$ spanning all of $j$ and $k$ space. Quantities are volume averaged from the 3D region to the 1D region to be used as boundary conditions for the 1D region; while values in the 1D region are copied across $j$ and $k$ space to the 3D region to be used as boundary conditions there. 

To date, only exploratory time trials have been performed because future additions to the code (including both an implicit solution to the energy equation with radiation diffusion and an eddy viscosity sub-grid-scale model) will impact the timing results significantly. At present, a calculation with 97 $\times$ 10 $\times$ 10 3D zones and 10 1D zones for 1 million time steps ( 10 million seconds, or approximately 178 fundamental mode periods) takes 2 processors 12$^h$22$^m$; 4 processors 5$^h$25$^m$; 8 processors $3^h28^m$; and 16 processors $2^h29^m$. At larger numbers of processors with the current gridding, the overhead from MPI begins to negate any additional benefits. Thus, for this gridding 16 processors represents the ``sweet spot''. At larger horizontal grid sizes the ``sweet spot'' will likely be pushed to larger numbers of processors. 

\section{TEST CASE RESULTS}

\begin{figure}[tb!]
\center
\plotone{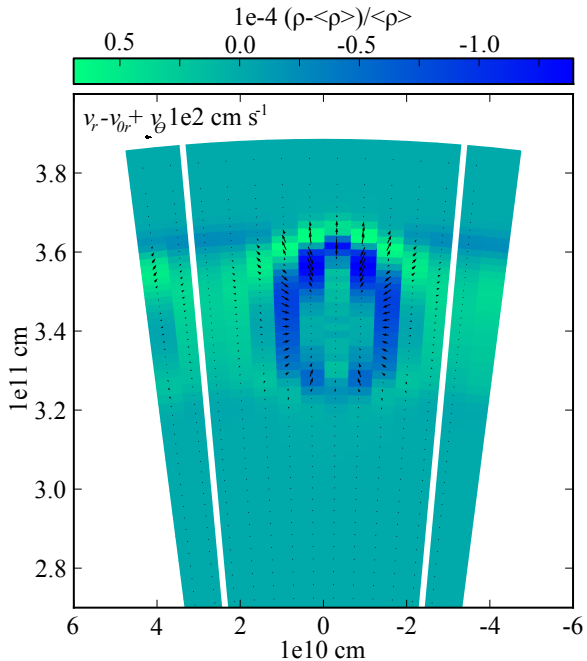}
\caption{This figure shows a two-dimensional slice at constant $\phi$ (3$^{\rm rd}$ $\phi$ zone) for CalIII. The color scale is $(\rho_{i,j,k}-\langle\rho\rangle_i)/\langle\rho\rangle_i)$ and vectors show the difference between the radial velocity and the grid velocity added vectorially with the $\theta$ velocity. The slice is at 7225 s into the calculation. Cells exterior to the white lines on the left and right sides are used to express the horizontal periodic boundary conditions. This figure shows only the outer 30\% of the stellar radius, while the total simulation is of more than 85\%.}
\label{fig:r-theta-slice-horizontal}
\end{figure}
\begin{figure}[tb!]
\center
\plotone{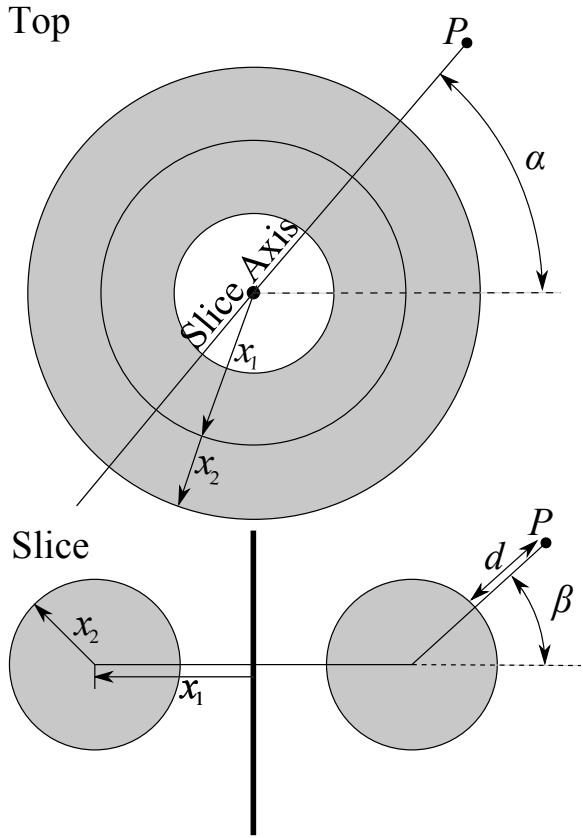}
\caption{This figure shows the geometry of the torus used to define the toroidal velocity perturbations. $x_1$ and $x_2$ are two radii used to define the equation of torus. $\alpha$ and $\beta$ are two angles used to define a point on the surface of the torus. The upper panel shows a top down view of the torus, while the lower panel shows the side view along the slice axis indicated in the top panel. The distance from an arbitrary point $P$ to the surface of the torus is given by $d$.}
\label{fig:torus-geo}
\end{figure}
\begin{figure}[tb!]
\center
\plotone{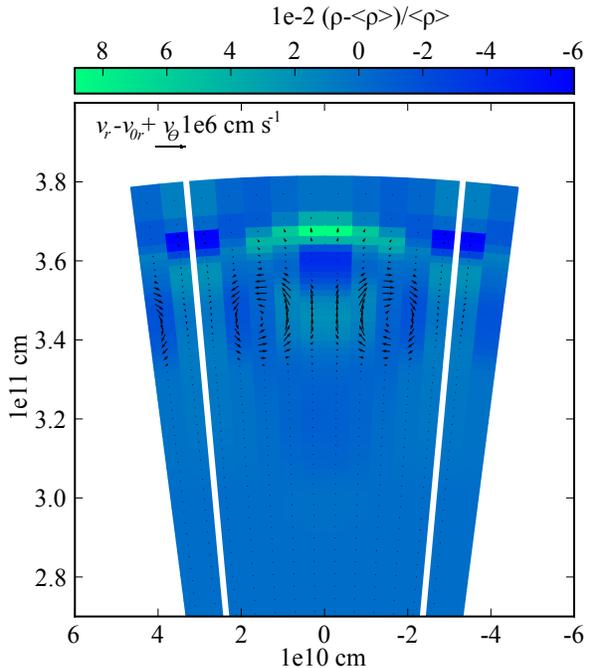}
\caption{This figure shows a two-dimensional slice at constant $\phi$ (6$^{\rm th}$ $\phi$ zone) with vectors showing the difference between the radial velocity and the grid velocity added vectorially with the $\theta$ velocity. This plot is from CalV at 9031 s into the calculation. At later times the initial toroidal velocity perturbation has spread through out the model, making its initial form indiscernible.}
\label{fig:r-theta-slice-torus}
\end{figure}
\begin{figure}[tb!]
\center
\plotone{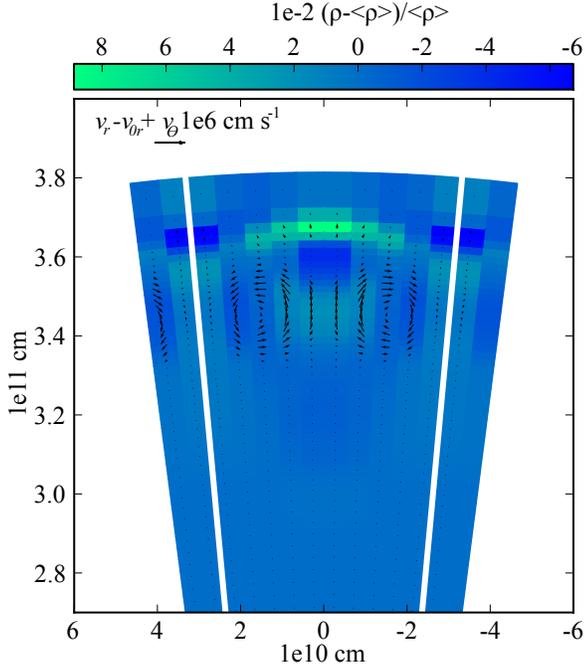}
\caption{This figure shows a two-dimensional slice at constant $\theta$ (6$^{\rm th}$ $\theta$ zone) with vectors showing the difference between the radial velocity and the grid velocity added vectorially with the $\phi$ velocity. For the same calculation and time as figure \ref{fig:r-theta-slice-torus}.}
\label{fig:r-phi-slice-torus}
\end{figure}
\begin{figure}[h!]
\center
\plotone{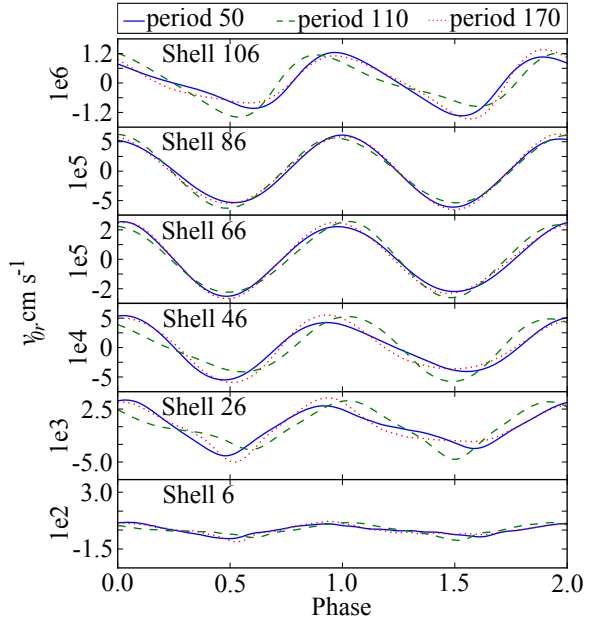}
\caption{This figure shows the radial grid velocity for three well separated periods, at six radial shells of the model for CalIV.}
\label{fig:reproducibility-u0}
\end{figure}
\begin{figure}[tb!]
\center
\plotone{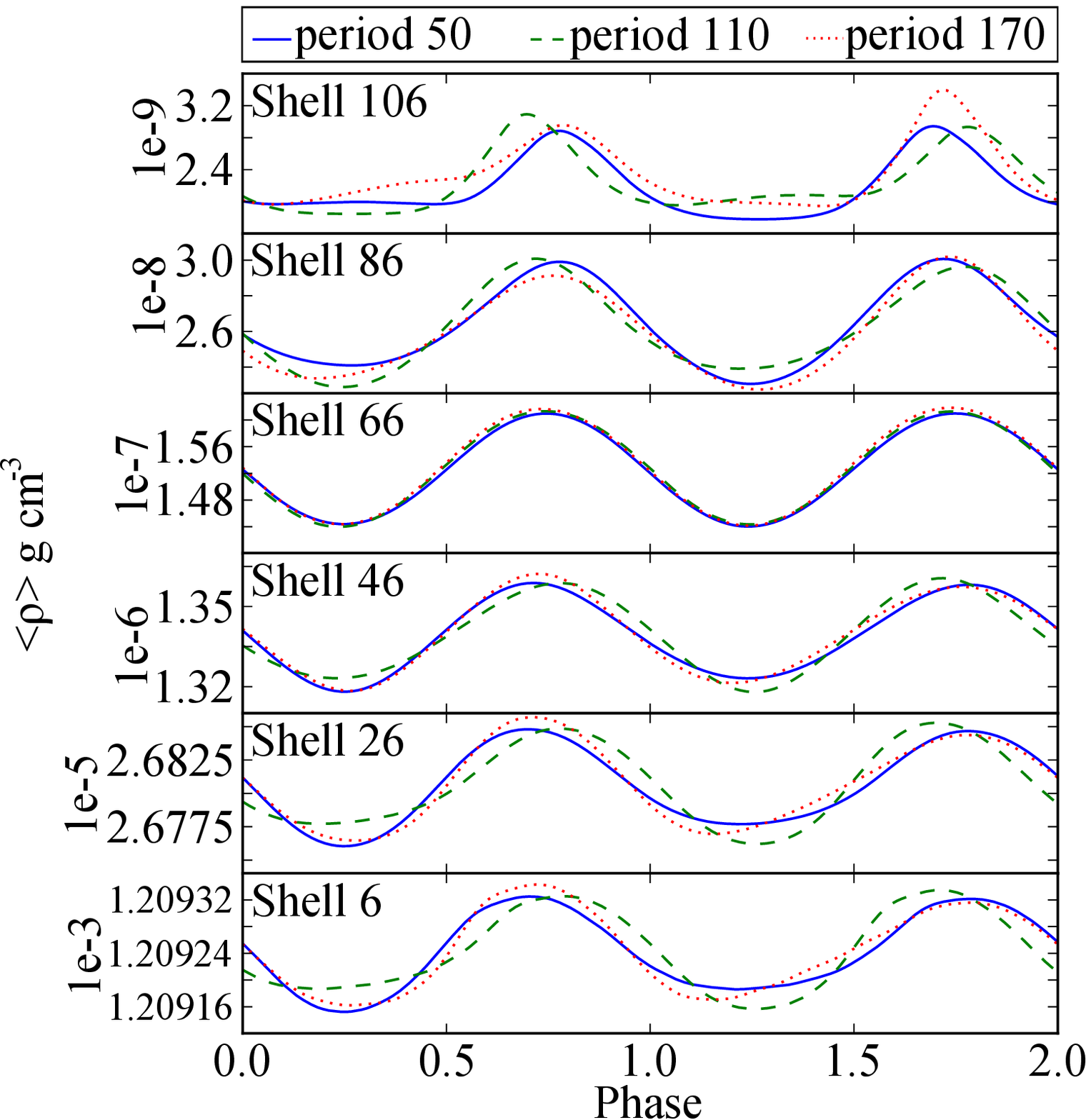}
\caption{This figure is similar to figure \ref{fig:reproducibility-u0} but for $\langle\rho\rangle$. The density is well reproduced across a large span of periods through the model, indicating that there is no drift of the radial coordinate system.}
\label{fig:reproducibility-DA}
\end{figure}
Five adiabatic test calculations have been performed using the method outlined above. The first calculation (CalI) was of a static stellar model with all velocities set to zero, which was used to test that the starting model was indeed generated in equilibrium consistently with respect to the hydrodynamic finite difference equations, and that SPHERLS's finite difference and finite volume representations of the hydrodynamic equations are hydodynamically stable. The second calculation (CalII) was a spherical blast wave \citep[e.g.][]{Sedov-1959}, used to test that the code could handle strong shocks and check it against an analytical solution. The third calculation (CalIII) was of a low amplitude radial pulsation (1 km s$^{-1}$ initial surface velocity) in the fundamental mode with a horizontal velocity perturbation to break spherical symmetry. The forth calculation (CalIV) was of an even lower amplitude radial pulsation (0.1 km s$^{-1}$ initial surface velocity) in the first overtone mode again with a horizontal velocity perturbation. These two low amplitude calculations were used to compare the calculated periods to the linear adiabatic periods, and to insure that the scheme worked as expected at low velocities. The lower velocity is needed for comparison with the linear LNA code which assumes small perturbations from the static model to calculate periods. The fifth calculation (CalV) started with the same stellar model as CalIII, but with a higher amplitude pulsation (10 km s$^{-1}$ initial surface velocity) and a toroidal velocity perturbation instead of the horizontal velocity perturbation. This calculation was used to show that SPHERLS reproduced quantities well from one period to the next over many periods in the presence of weak shocks and large scale structured motions.

The static calculation (CalI) started with all velocities ($r$, $\theta$, and $\phi$) set to zero and was computed for over 20 million seconds (356 fundamental periods). The radial velocities in the surface zone reached the largest amplitudes. In this zone the radial velocity amplitude initially grew from zero to a temporal mean of about $3.7\times 10^{-5}$ cm s$^{-1}$ within the first 80 periods. This mean was maintained for the rest of the calculation with a standard deviation of $2.7\times 10^{-5}$ cm s$^{-1}$. The horizontal velocities remain zero throughout the calculation. This is understandable because all terms in the horizontal momentum equations are zero and remain that way; while the radial momentum equation is the balance between two nonzero terms and thus subject to round off error.

To assure that the method is behaving as designed we checked the mass calculated from the averaged density, $\langle\rho\rangle$, and the shell volume (which is dependent on the radial grid velocity) with the mass set as the independent variable. The largest relative difference between the calculated mass and the independent mass variable for all calculations is $4\times 10^{-13}$. This is only two significant digits above machine round off, and there are no signs of a trend with time.

To test how well SPHERLS handles strong shocks and to compare the computed results with an analytic calculation we performed a Sedov blast wave calculation (CalII). This calculation had 400 radial zones with an initial spacing of 10 cm and 10 $\theta$ and $\phi$ zones with spacings of 1$^{\circ}$. The 400 radial zones produces a 40 m radius spherical volume for the shock to expand into, and was chosen to allow enough volume for the shock to expand into over 10 ms, at which time the analytic solution has reached a shock radius of 32.7 m. The inner 10 radial zones were (as in the adiabatic stellar models) treated in 1D. The blast was accomplished by setting the initial energy in the inner 30 zones (i.e., a 3 m radius sphere) to $4.18 \times 10^{21}$ ergs with all other zones having an energy of $1\times 10^6$ ergs. The density was set to 2 g cm$^{-3}$ through out the starting model. A gamma-law gas was used for the equation of state, with a $\gamma$ of 1.6. All the initial velocities were set to zero and the blast was followed for 10 ms. The calculation was compared to an analytical solution with a point-source energy producing the blast, evaluated at times from 0.5 ms to 10 ms in 0.5 ms intervals. When comparing the extended-source to the point-source solution one would expect that at early times (when the shock is closer to the source), and at later times closer to the initial location of the source, the discrepancies between the computed and analytical solutions should be larger. This is because the differences between a non-point source calculation and the point source analytic solution will diminish as the disturbance moves outward. The blast radii computed by SPHERLS matched those form the analytic solution to within 7.5 cm at all times. The best match of shock radii (within 1.7 cm) occurred later in the calculation at a time of 10 ms, while the worst match (7.3 cm ) occurred much earlier in the calculation at 3.5 ms.  The computed velocity, density, and pressure profiles were also compared to the analytic solution, but because of the extended source, only the zones outside the initial explosion source were compared. The root mean square of the fractional error in velocity, density, and pressure was less than 3\%, 8\%, and 5\% respectively in the last half of the calculation (5 ms to 10 ms). In the first half of the calculation (0.5 ms to 5 ms) the root mean square of the fractional errors are a bit larger, mostly due to the difference between using an extended source in the calculation versus a point source in the analytic solution and are within 8\%, 15\%, and 11\% for the velocity, density, and pressure respectively. The radial profiles of the velocity, density, and pressure fit quite well without any outlying points.

Because the calculations are adiabatic, we expect the pulsation to neither grow nor decay and to be reproducible from one period to the next. This should provide a good test to verify that our numerical algorithm functions as desired over many periods. Both the low amplitude fundamental and first overtone pulsation (CalIII and CalIV respectively) had a horizontal velocity perturbation imposed on them to break spherical symmetry, by setting specific values of $v_\theta$ and $v_\phi$ at a central horizontal zone located at 90\% of the total radius (18 zones in from the surface of the 107 radial zone models). The velocities were directed horizontally out of the zone through sides $A_{j\pm 1/2}$ and $A_{k\pm 1/2}$ (see figure \ref{fig:cell}). The magnitude of these horizontal velocities was taken to be half of the initial radial velocity at this radial location (0.3 km s$^{-1}$ and 0.03 km s$^{-1}$ for CalIII and CalIV respectively). Figure \ref{fig:r-theta-slice-horizontal} shows a two-dimensional slice at constant $\phi$ of CalIII slightly after the initial conditions. The slice is at 7225 s into the calculation (relatively early in the $1 \times 10^7$ s calculation) and shows the disturbance resulting from the horizontal velocity perturbation as well as it's location and geometry with respect to the rest of the model.

The period of the fundamental mode is 56178 s for the low amplitude calculation (CalIII), and compares well with the period calculated from LNA of 56114 s. There is less than $0.12$\% difference between the periods of the two codes. The first overtone model (CalIV) was found to have a period of 38911 s and compares with the LNA period of 39522 s, producing less than a 1.6\% difference.

In addition to the horizontal velocity perturbation we explored in CalIII and CalIV we also explored a velocity perturbation that is more structured over a larger scale (CalV). To create this model we started with the same structural model and radial velocity profile as Cal III, this time however, using a surface amplitude of 10 km s$^{-1}$. On top of the radial velocity profile we added a velocity perturbation in the shape of a torus (see figure \ref{fig:torus-geo} for torus geometry). The velocity perturbations were taken to be constant on the surface of the torus (the two circles in the lower half of figure \ref{fig:torus-geo}) and parallel to the surface of the torus. By locating the closest point (defined by angles $\alpha$ and $\beta$) on the surface of the torus to the point $P$, the distance $d$ in figure \ref{fig:torus-geo} can be calculated. Then a Gaussian centered on the surface of the torus with a maximum amplitude of 5 km s$^{-1}$ is evaluated at $d$ providing the velocity magnitude. The FWHM of the Gaussian is chosen so that the velocity perturbations do not overlap the other parts of the torus, and so that the velocity perturbations are still reasonably strong a zone or two away from the surface of the torus. The direction of the velocity is taken to be parallel to the surface of the torus at the locaiton closest to $P$. The velocity magnitude is then broken into $r$, $\theta$, and $\phi$ components. The result of applying this perturbation is shown in figures \ref{fig:r-theta-slice-torus} and \ref{fig:r-phi-slice-torus}. These figures show slices through the center of the model at constant $\theta$ and $\phi$ respectively, 9031 s into the calculation and indicate that $\theta$ and $\phi$ directions behave identically.

The high surface velocity model (CalV) results are presented in figures \ref{fig:reproducibility-u0} and \ref{fig:reproducibility-DA}. Figure \ref{fig:reproducibility-u0} shows that the radial grid velocity throughout the model is well reproduced at periods 50, 110, and 170. Figure \ref{fig:reproducibility-DA} shows that $\langle\rho\rangle$ is well reproduced over a large number of periods and does not show the drifting apparent in \citep{Deupree-1977a}. The fact that the calculations of dependent quantities are reproduced over many periods shows that the scheme is working as desired and we expect to calculate full amplitude solutions of pulsating variable stars in the future.

A low radial surface velocity model(1 km s$^{-1}$ surface velocity) that was spherically symmetric and did not use artificial viscosity has a very constant peak kinetic energies per period, with out any long term detectable growth or decay rates larger than $1\times 10^{-11}$\% per fundamental period. While CalIII, that had a horizontal velocity perturbation, had a growth rate of $4.3\times 10^{-4}$\% per fundamental period. In the high radial surface velocity calculation (CalV) artificial viscosity is required in the momentum equations (eq. \ref{eq:rad-mom-cons}, \ref{eq:theta-mom-cons}, and \ref{eq:phi-mom-cons}) to reduce the very steep pressure gradients at shocks.  If it is omitted when pulsation amplitudes exceed the local sound speed it ultimately leads to negative densities and energies. One might argue that including the artificial viscosity in the energy equation would produce a non-adiabatic calculation, since the inclusion of the artificial viscosity raises the internal energy more than otherwise when the volume is decreasing. In CalV if the artificial viscosity is included in the energy equation we find that the peak kinetic energy decays at a rate 0.36\% per fundamental period. If it isn't included the peak kinetic energy decays at a rate of 0.13\%. These decay rates are dependent on the inclusion or omission of artificial viscosity in the energy equation, and may affect the amplitude of full amplitude solutions to some degree. This is not merely a concern for the present code, but for all non-linear hydrodynamics codes that use artificial viscosity.

\section{CONCLUSIONS AND NEXT STEPS}
We have developed a numerical algorithm and the computer code, SPHERLS, which is able to follow 3D adiabatic radial pulsations for many periods when spherical symmetry is broken. This is a necessary step for following the convective motions and radial pulsations of stars. We have shown that SPHERLS maintains hydrostatic equilibrium to a high degree over 356 fundamental periods. The radial Lagrangian algorithm for maintaining constant mass in the radial shells is effective over 178 periods to within one or two digits of machine round off, with no signs of any particular trend either decreasing or increasing.

We have found that SPHERLS reproduces both the fundamental and first overtone modes of the linear adiabatic code LNA reasonably well ($<$0.12\% for fundamental and $<$ 1.6\% for first over tone). The velocities as well as other dependent quantities (e.g. the horizontally averaged density) are reproducible over many periods when spherical symmetry is broken, again indicating that our radial Lagrangian scheme is performing as designed.

Adiabaticity is maintained extremely well at low radial velocities when artificial viscosity is not included. However at higher radial velocities some kinetic energy is converted into internal energy via the artificial viscosity required to smooth out shocks. This should be kept in mind while using any non-linear hydrodynamics code employing artificial viscosity to compare full amplitude solutions with observations.

In order to calculate full amplitude pulsating models to compare with observations a few additions must be made to SPHERLS. A more realistic equation of state and radiative Rosseland mean opacities must be included and radiation diffusion must be added to the energy equation. The latter is expected to require an implicit integration of the energy equation, at least near the surface, because optically thin zones would require a very small time step based on the speed of light and not the speed of sound. We expect to use the current explicit solution to the energy equation deeper in the envelope where the mean free path of photons is much less than a computational zone, keeping the calculation time down. Finally, we will add the subgrid scale terms required in a large eddy simulation for treating turbulent convection in the ionization zone.

\acknowledgments

These calculations were performed with ACEnet computational resources. ACEnet, a part of Compute Canada, provides academic high performance computing to Atlantic Canada. CMG is supported in part by an NSERC Discovery Grant to RGD and in part by an ACEnet fellowship.


\end{document}